\documentclass[aps,twocolumn,superscriptaddress,showpacs]{revtex4}

\usepackage{psfig,epsfig}
\usepackage{latexsym}
\usepackage{graphicx}
\usepackage{amsmath}
\usepackage{pstricks}
\usepackage{pst-node,pst-plot}
\begin{document}


\bibliographystyle{apsrev}

\title{Quantum Brain: A Recurrent Quantum Neural Network Model to
  Describe Eye Tracking of Moving Targets}

\author{ Laxmidhar Behera} \affiliation{Department of Electrical Engineering,
Indian Institute of Technology, Kanpur
208 016, UP, INDIA}

 \author{Indrani Kar} \affiliation{Department of Electrical Engineering,
Indian Institute of Technology, Kanpur
208 016, UP, INDIA}

\author{Avshalom Elitzur} \affiliation{Unit of Interdisciplinary Studies,
  Bar-IIan University, 52900 Ramat-Gan, Israel}

\begin{abstract}
A theoretical quantum brain model is proposed using a nonlinear Schroedinger
wave equation. The model proposes that there exists a quantum process that
mediates the collective response of a neural lattice (classical brain). The model is used to explain eye movements when tracking
moving targets.  Using
a Recurrent Quantum Neural Network(RQNN) while simulating the quantum brain model, two very
interesting phenomena are observed. First, as eye sensor data is processed in a classical brain, a wave packet is triggered in the quantum
brain. This wave packet moves like a particle. Second, when the eye tracks a fixed target, this wave packet moves not in a continuous but rather in a discrete mode. This result reminds one of the saccadic movements of the eye consisting of 'jumps' and 'rests'.  However, such a saccadic movement is intertwined with smooth pursuit movements when the eye has to track a dynamic trajectory. In a sense, this is the first theoretical model explaining the
experimental observation reported concerning eye movements in a static scene
situation. The resulting prediction is found to be very precise and efficient in comparison to classical objective modeling
schemes such as the Kalman filter. 
\end{abstract}

\maketitle

\section{Introduction}
Information processing in the brain is mediated by the dynamics of large, highly interconnected neuronal populations. The activity patterns exhibited by the brain are extremely rich; they include
stochastic weakly correlated local firing, synchronized oscillations and
bursts, as well as propagating waves of activity. Perception, emotion etc. are supposed to be emergent properties of such a complex nonlinear neural circuit. 

Instead of considering one of the conventional neural
architectures \citep{Cohen:83,Amari:83,Behera:96,Behera:98,Amit:89}, an alternative neural
architecture  is proposed here for neural computing. Indeed, there are certain
aspects of brain functions that still appear to have no satisfactory
explanation.  As an alternative, researchers \citep{Tuszynski:95, Vitiello:95, Hagan:00, Mershin:00} are investigating whether the brain can demonstrate quantum mechanical behavior. According to current research, microtubules, the basic
components of neural cytoskeleton, are very likely to possess quantum
mechanical properties due to their size and structure. The tubulin protein, which is the structural block of microtubules, has the ability to flip from one conformation to another as a
result of a shift in the electron density localization from one resonace
orbital to another. These two conformations act as two basis states of the
system according to whether the electrons inside the tubuline hydrophobic
pocket are localized closer to $\alpha$ or $\beta$ tubulin. Moreover the
system can lie in a superposition of these two basis states, that is, being in
both states simultaneously, which can give a plausible mechanism for creating
a coherent state in the brain. To give credence to the possibility of
existence of a quantum brain, Penrose (\citep{Penrose:94}) argued that the human brain must utilize quantum mechanical effects when
demonstrating problem solving feats that cannot be explained algorithmically.

In this paper, instead of going into biological details of the brain, we propose a theoretical quantum brain model. The model is
referred to as Recurrent Quantum Neural Network (RQNN). An earlier version \citep{Behera:03} of this model used a linear neural circuit to set up the potential field in which the quantum brain is dynamically excited. The present model
uses a nonlinear neural circuit. This fundamental change in the architecture
has yielded two novel features. The wave packets, $f(x,t)=\mid\psi(x,t)\mid^2$, are moving like 
particles. Here $\psi(x,t)$ is the solution of the nonlinear Schroedinger wave equation that describes the quantum brain model proposed in this paper to
explain eye movements for tracking moving targets. The other very interesting
observation is that the movements of the wave packets while tracking a fixed
target are not continuous but discrete. These observations accord with the well-known saccadic movement of the eye \citep{Bahill:79}. In a way, our model is the first of its kind to explain the nature of eye movements in static scenes that consists of "jumps" (saccades) and "rests" (fixations). We expect this result to inspire other researchers to further investigate the possible quantum dynamics of the brain.

\section{A Theoretical Quantum Brain Model}
\label{sec:qbt}
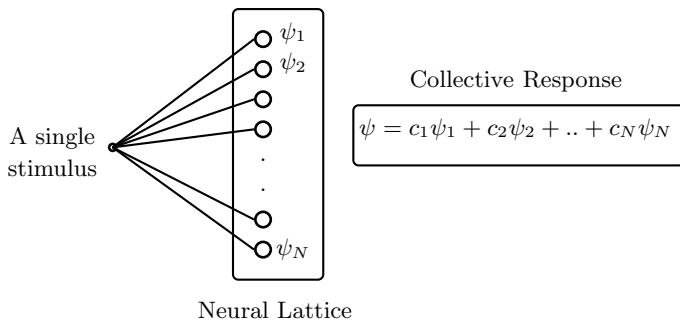
\begin{figure}[htbp]
\begin{center}\psset{unit=0.8cm}
    \begin{pspicture}(12,6)

      \pscircle[linewidth=1pt](2.0,3.7){0.08}
      \pscircle[linewidth=1pt](4.5,4.0){0.15}
      \pscircle[linewidth=1pt](4.5,4.5){0.15}
      \pscircle[linewidth=1pt](4.5,5.0){0.15}
      \pscircle[linewidth=1pt](4.5,5.5){0.15}
      \pscircle[linewidth=1pt](4.5,2.0){0.15}
      \pscircle[linewidth=1pt](4.5,2.5){0.15}

      \rput(4.5,3.0){$.$}
      \rput(4.5,3.5){$.$}

      \psline{-}(2.0,3.7)(4.35,4.0)
      \psline{-}(2.0,3.7)(4.35,4.5)
      \psline{-}(2.0,3.7)(4.35,5.0)
      \psline{-}(2.0,3.7)(4.35,5.5)
      \psline{-}(2.0,3.7)(4.35,2.5)
      \psline{-}(2.0,3.7)(4.35,2.0)

      \rput(5,5.6){$\psi_1$}
      \rput(5,5.1){$\psi_2$}
      \rput(5,2.0){$\psi_N$}
      
      \pspolygon[linearc=2pt](4.0,1.5)(5.5,1.5)(5.5,6)(4.0,6)
      \rput(4.7,1){\small Neural Lattice}
      \rput(1.0,3.85){\small A single}
      \rput(1.0,3.35){\small  stimulus}

      \pspolygon[linearc=2pt](6.0,3.4)(6.0,4.4)(11.5,4.4)(11.5,3.4)
      \rput(8.7,4){\begin{tabular}{lll} {\small $\psi = c_1 \psi_1 + c_2 \psi_2 + .. +c_N \psi_N$}\end{tabular}}
      
      \rput(8.7,4.8){\small Collective Response}
 
    \end{pspicture}
  \end{center}
  \caption{Quantum Brain - A Theoretical Model}
  \label{fig:qb}
\end{figure}

An impetus to hypothesize a quantum brain model comes
from the brain's necessity to unify the neuronal response into a single percept. Anatomical, 
neurophysiological and neuropsychological evidence, as well as brain imaging
using fMRI and PET scans, show that separate functional maps exist in the biological
brain to code separate features such as direction of motion, location, color and orientation. How does the brain compute all this data to have a coherent perception? In this paper, a
very simple model of a quantum brain is proposed where a collective response of a
neuronal lattice is modeled using a Schroedinger wave equation as shown in
FIG. \ref{fig:qb}. In this figure, it is shown that an external stimulus
reaches each neuron in a lattice with a probability amplitude function
$\psi_i$. Such a hypothesis would suggest that the carrier of the stimulus
performs quantum computation. The collective response of all the neurons is given by the superposition equation:

\begin{equation}
\psi = c_1 \psi_1 + c_2 \psi_2 + .. +c_N \psi_N=\sum_{i=0}^N c_i \psi_i
\end{equation}

We suggest that the time evolution of the collective response $\psi$ is described by the
Schroedinger wave equation:

\begin{equation}
\label{eq:swave}
i\hbar \frac{\partial \psi(x,t)}{\partial t}= - \frac{\hbar^2}{2m} \nabla^2
\psi(x,t) + V(x) \psi(x,t)
\end{equation}
where $2\pi\hbar$ is Planck's constant, $\psi(x,t)$ is the wave
function (probability amplitude) associated with the quantum object at space-time point$(x,t)$, and $m$ the mass of the quantum object. Further symbols such as $i$ and $\nabla$ carry their usual meaning in the context of the Schroedinger wave equation. Another way to look at our proposed quantum brain is as follows. A neuronal lattice sets up a spatial potential
field $V(x)$. A quantum process described by a quantum state $\psi$ which
mediates the collective response of a neuronal lattice evolves in the spatial
potential field $V(x)$ according to equation \eqref{eq:swave}. Thus the
classical brain sets up a spatio-temporal potential field and the quantum
brain is excited by this potential field to provide a collective response. In 
the next section, we present a possible eye-movement model for tracking moving
target.

\section{An Eye Tracking Model}
\label{sec:rqnn_arch}

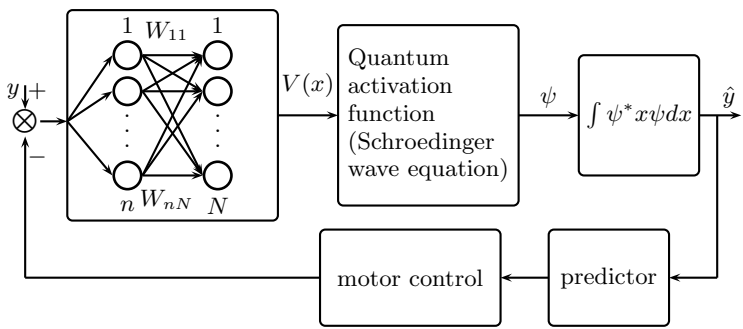
\begin{figure}[htbp]
  \begin{center}\psset{unit=0.8cm}
    \begin{pspicture}(12,6)
      \psline{->}(-0.45,3.6)(0,3.6)
      \rput(-0.9,4.1){$y$}
      \psline{->}(-0.7,4.2)(-0.7,3.8)
     \psline{->}(0,3.6)(0.75,4.7)
     \psline{->}(0,3.6)(0.75,2.7)
     \psline{->}(0,3.6)(0.75,4.1)
      \psline{->}(1.25,4.7)(2.25,4.7)
      \psline{->}(1.25,4.7)(2.25,2.7)
      \psline{->}(1.25,4.1)(2.25,4.7)
      \psline{->}(1.25,4.1)(2.25,2.7)
      \psline{->}(1.25,2.7)(2.25,4.7)
      \psline{->}(1.25,2.7)(2.25,4.1)
      \psline{->}(1.25,2.7)(2.25,2.7)
      \rput(1.0,5.2){$1$}
      \rput(2.5,5.2){$1$}
      \rput(1.65,5.1){$W_{11}$}
      \rput(1.65,2.3){$W_{nN}$}
      \rput(1,2.2){$n$}
      \rput(2.5,2.2){$N$}

      \pspolygon[linearc=2pt](0,5.45)(3.5,5.45)(3.5,1.95)(0,1.95)
      \psline{->}(3.5,3.7)(4.5,3.7)
      \rput(4,4.2){$V(x)$}
      \pscircle[linewidth=1pt](2.5,4.7){0.25}
      \pscircle[linewidth=1pt](2.5,4.1){0.25}
      \rput(2.5,3.7){$.$}
      \rput(2.5,3.2){$.$}
      \rput(-0.7,3.6){$\bigotimes$}
      \rput(2.5,3.45){$.$}
      \rput(1.0,3.7){$.$}
      \rput(1.0,3.2){$.$}
      \rput(1.0,3.45){$.$}
      \pscircle[linewidth=1pt](1.0,4.7){0.25}
      \pscircle[linewidth=1pt](1.0,4.1){0.25}
      \pscircle[linewidth=1pt](1.0,2.7){0.25}
      \psline{->}(1.25,4.7)(2.25,4.1)
      \psline{->}(1.25,4.1)(2.25,4.1)
      \pscircle[linewidth=1pt](2.5,2.7){0.25}
      \pspolygon[linearc=2pt] (4.2,1.8)(7.2,1.8)(7.2,0.2)(4.2,0.2)
      \rput(5.7,1){motor control}
      \pspolygon[linearc=2pt] (8.0,1.8)(10.0,1.8)(10.0,0.2)(8.0,0.2)
      \rput(9.0,1){predictor}
      \pspolygon[linearc=2pt] (4.5,5.2)(7.5,5.2)(7.5,2.2)(4.5,2.2)
      \rput(6,3.7){\begin{tabular}{lll} Quantum\\  activation\\
       function\\ (Schroedinger \\ wave equation)\end{tabular}}
   \psline{->}(7.5,3.7)(8.5,3.7)
      \pspolygon[linearc=2pt] (8.5,4.7)(10.5,4.7)(10.5,2.7)(8.5,2.7)
      \rput(9.5,3.7){$\int \psi^{*} {x} \psi dx$}
      \rput(8,4){$\psi$}
      \psline{->}(10.5,3.7)(11.2,3.7)
      \psline{-}(10.8,3.7)(10.8,1.0)
      \psline{->}(10.8,1.0)(10,1.0)
      \psline{->}(8,1.0)(7.2,1.0)
      \psline{-}(4.2,1.0)(-0.7,1.0)
      \psline{->}(-0.7,1.0)(-0.7,3.35)
      \rput(-0.5,3.0){$-$}
      \rput(-0.5,4.1){$+$}
      \rput(11,4){$\hat y$}
    \end{pspicture}
  \end{center}
  \caption{Conceptual framework for the Recurrent Quantum Neural Networks}
  \label{fig:qnn}
\end{figure}

Let us consider a plausible biological mechanism for eye tracking using
the quantum brain model proposed in section \ref{sec:qbt}. The mechanism 
of eye movements tracking a moving target consists of three stages as shown in
FIG. \ref{fig:qnn}: (i) stochastic filtering of noisy data that impact the eye
sensors; (ii) a predictor that predicts the next spatial position of the
moving target; and (iii) a biological motor control system that aligns the eye pupil along the moving targets trajectory. 
The biological eye sensor fans out the input signal $y$ to a specific neural lattice in the visual cortex. For clarity, Figure \ref{fig:qnn} shows a one-dimensional array of neurons whose receptive fields are excited by the signal input $y$ reaching each neuron through a synaptic connection described by a
nonlinear map. The neural lattice responds to the stimulus by setting up a
spatial potential field, $V(x,t)$, which is a function of external stimulus $y$ and estimated trajectory $\hat{y}$ of the moving target:
 
\begin{equation}
\label{eq:pot}
V(x,t)=\sum_{i=1}^n W_i(x,t) \phi_i(\nu(t))
\end{equation}
where $\phi_i(.)$ is a Gaussian Kernel function, $n$ represents the number of
such Gaussian functions describing the nonlinear map that represents the synaptic connections, $\nu(t)$
represents the difference between $y$ and $\hat{y}$ and $W$ represents the synaptic weights as shown in FIG. \ref{fig:qnn}. The
Gaussian kernel function is taken as:

\begin{equation}
 \phi_i(\nu(t))= exp(-(\nu(t)-g_i)^2)
\end{equation}
where $g_i$ is the center of the $i^{th}$ Gaussian function, $\phi_i$. This center
is chosen from input space described by the input signal, $\nu(t)$, through uniform random sampling.

Our quantum brain model proposes that a quantum process mediates the collective
response of this neuronal lattice which sets up a spatial potential field $V(x,t)$. This happens when the quantum state
associated with this quantum process evolves in this potential field. The
spatio-temporal evolution follows as per equation
\eqref{eq:swave}. \textit{We hypothesize that this collective response is
  described by a wave packet, $f(x,t)=\mid \psi(x,t)\mid^2$, where the term
  $\psi(x,t)$ represents a quantum state.} In a generic sense, we assume that a
classical stimulus in a brain triggers a wave packet in the  counterpart
'\textit{quantum brain}'. This subjective response, $f(x,t)$, is quantified
using the following estimate equation:

\begin{equation}
\hat y(t)=\int x(t) f(x,t) dx
\end{equation}
 
The estimate equation is motivated by the fact that the wave packet,
$f(x,t)=\mid\psi(x,t)\mid^2$ is interpreted as the probability density function. Based on this estimate,
$\hat{y}$, the predictor estimates the next spatial position of the moving
target. To simplify our analysis, the predictor is made silent. Thus its
output is the same as that of $\hat{y}$. The biological motor control is commanded
to fixate the eye pupil to align with the target position, which is predicted
to be at $\hat{y}$. Obviously, we have assumed that biological motor control is ideal.

After the above mentioned simplification, the closed form dynamics of the model described by Figure \ref{fig:qnn} becomes:

\begin{equation}
\label{eq:wave1}
\begin{aligned}
i\hbar \frac{\partial \psi(x,t)}{\partial t} = - \frac{\hbar^2}{2m} \nabla^2
\psi(x,t)+\;\;\;\;\;\;\;\;\;\;\;\;\;\;\;\;\;\;\;\;\;\;\;\;\;\;\;\;\;\;\;\;&\\\zeta G \left(y(t) - \int x \mid \psi(x,t) \mid^2 dx \right) \psi(x,t)
\end{aligned}
\end{equation}
where $G(.)$ is a Gaussian kernel map
introduced to nonlinearly modulate the spatial potential field that excites
the dynamics of the quantum object. In fact $\zeta G(.)=V(x,t)$ where $V(x,t)$
is given in equation \eqref{eq:pot}.

The nonlinear Schroedinger wave equation given by equation \eqref{eq:wave1} is
one-dimensional with cubic nonlinearity. Interestingly, the closed form
dynamics of the Recurrent Quantum Neural Network (equation
\eqref{eq:wave1}) closely resembles a
nonlinear Schroedinger wave equation with cubic nonlinearity studied in quantum electrodynamics \cite{gupta:01}:

\begin{equation}
\label{eq:nse}
\begin{aligned}
i\hbar\frac{\partial \psi(x,t)}{\partial t}=\left(-\frac{\hbar^2}{2m}
  \nabla^2-\frac{e^2}{r}\right) \psi(x,t) +\\e^2\int \frac{\psi(x,t) \mid
\psi(x',t)\mid^2}{\mid x-x'\mid} dx'
\end{aligned}
\end{equation}
where $m$ is the electron mass, $e$ the elementary charge and $r$
the magnitude of $\mid x\mid$. Also, nonlinear Schroedinger wave equations with cubic nonlinearity
of the form $ \frac{\partial}{\partial
  t}\mathcal{A}(t)=c_1\mathcal{A}+c_3\mid \mathcal{A} \mid^2 \mathcal{A}$,
where $c_1$ and $c_3$ are constants, frequently appear in
nonlinear optics \cite{ Boyd:91}  and in the study of
solitons \cite{Jackson:91, Birula:76, Davydov:82, Scott:73}.

In equation \eqref{eq:wave1}, the unknown parameters are weights $W_i(x,t)$
associated with the Gaussian kernel, mass $m$, and $\zeta$, the scaling factor to
actuate the spatial potential field. The weights are updated using the Hebbian learning algorithm

\begin{equation}
\label{eq:beta}
\frac{\partial W_i(x,t)}{\partial t}=\beta \phi_i(\nu(t)) f(x,t)
\end{equation}
where $\nu(t)=y(t)-\hat{y}(t)$.

The idea behind the proposed quantum computing model is as
follows. As an individual
observes a moving target, the uncertian spatial position of the moving target
triggers a wave packet within the quantum brain. The quantum brain is so
hypothesized that this wave packet turns out to be
a collective response of a classical neural lattice. As we combine equations \eqref{eq:wave1}
and \eqref{eq:beta}, it is desired that there exist some parameters $m$, $\zeta$
and $\beta$ such that each specific spatial position $x(t)$ triggers a unique
wave packet, $f(x,t)=\mid \psi(x,t) \mid^2$, in the quantum brain. This brings
us to the question whether the closed form dynamics can exhibit soliton properties. As
pointed out above, our equation has a form that is known to possess soliton properties for
a certain range of parameters and we
just have to find those parameters for each specific problem. 

We would like to reiterate the importance of the soliton
properties. According to our model, eye tracking means tracking of a wave
packet in the domain of the quantum brain. The biological motor control aligns the eye pupil along the spatial position of the external target that the eye tracks. As the eye sensor receives data $y$ from this position, the resulting error stimulates the quantum brain. In a noisy background, if the tracking is accurate,
 then this error correcting signal $\nu(t)$ has little effect on the movement
of the wave packet. Precisely, it is the actual signal content in the input
$y(t)$ that moves the wave packet along the desired direction which, in effect,
achieves the goal of the stochastic filtering part of the eye movement for
tracking purposes. 

\section{Simulation Results}
In this section we present simulation results to test target tracking through
eye movement where targets are either fixed or moving.

\begin{figure}[htbp]
  \begin{center}
    \includegraphics[width=6.0cm,angle=0]{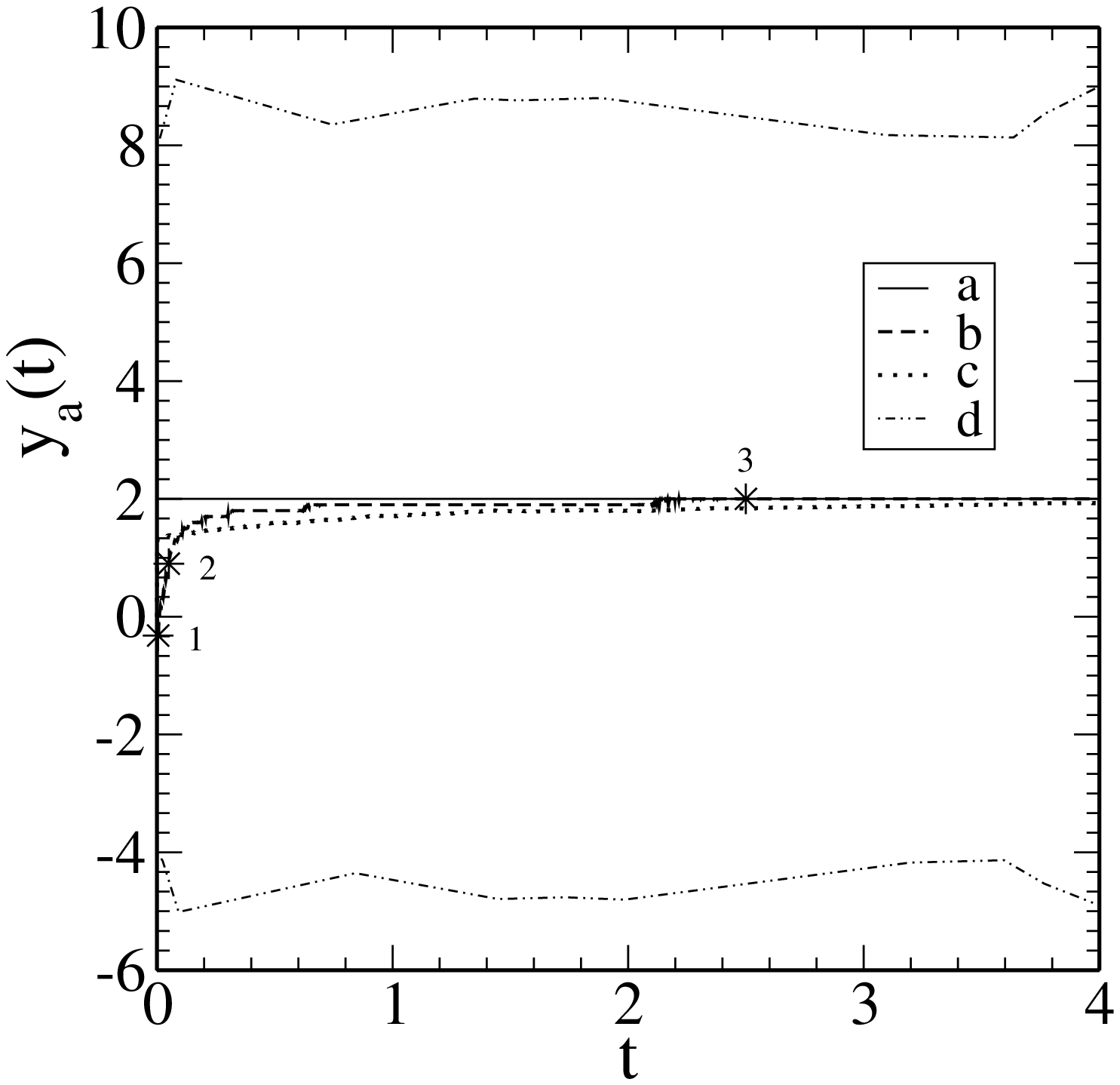}
    \includegraphics[width=6.0cm,angle=0]{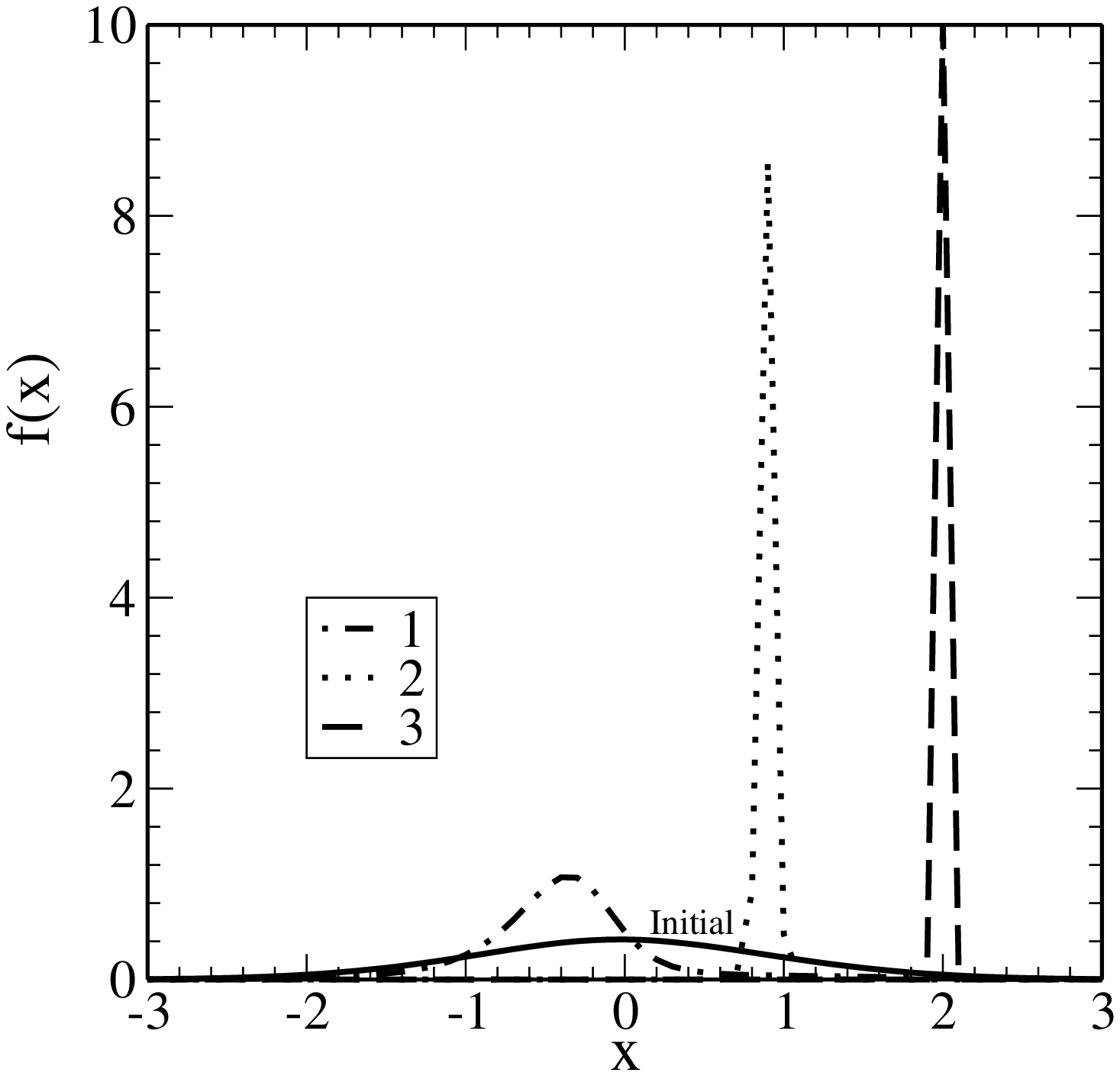}
    \caption{\small(\textit{top}) Eye tracking of a fixed target in a noisy
      environment of 0 dB SNR: 'a' respresents fixed target, 'b' represents
      target tracking using RQNN model and 'c' represents
      target tracking using a Kalman filter. The noise envelope is
      represented by the curve 'd'; (\textit{bottom})  The snapshots of the
      wave packets at different instances corresponding to the marker points
      (1,2,3) as shown in the \textit{top}
      figure. The solid line represent
      the initial wave packet assigned to the Schroedinger wave equation.} 
    \label{fig:dc_track3}
  \end{center}
\end{figure}

\begin{figure}[htbp]
  \begin{center}
    \includegraphics[width=6.0cm,angle=0]{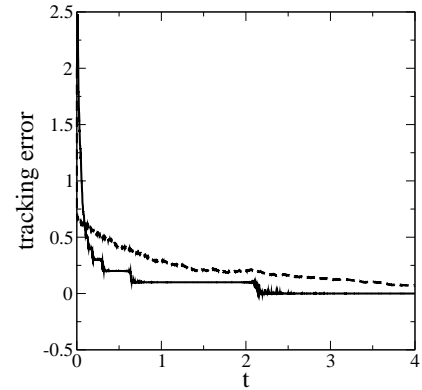}
    \caption{\small The continuous line represents tracking error using RQNN model
      while the broken line represents tracking error using Kalman filter} 
    \label{fig:error}
  \end{center}
\end{figure}

\begin{figure}[htbp]
  \begin{center}
    \includegraphics[width=6.0cm,angle=0]{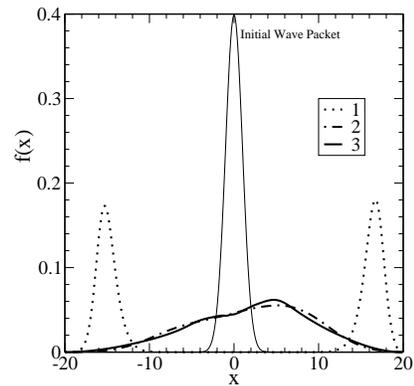}
    \caption{\small Wave packet movements for RQNN with linear weights } 
    \label{fig:ln_rqnn}
  \end{center}
\end{figure}

For fixed target tracking, we have
simulated a stochastic filtering problem of a dc signal embedded in Gaussian
noise. As the eye tracks a fixed target, the corresponding dc signal is taken as
$y_a(t)=2.0$, embedded in Gaussian noise with SNR (\textit{signal to noise ratio})values of 20 dB, 6 dB and 0 dB. 


We compare the results with the performance of a Kalman filter \citep{Grewal:01} designed for
this purpose. It should be noted that the Kalman filter has the \textit{a
  priori} knowledge that the embedded signal is a dc signal whereas the RQNN
is not provided with this knowledge. The Kalman filter also makes use of the
fact that the noise is Gaussian and estimates the variance of the noise based
on this assumption. Thus it is expected that the  performance of the Kalman
filter will degrade as the noise becomes non-Gaussian. In contrast, the RQNN model does
not make any assumption about the noise. 

It is observed that there are certain values of $\beta$, $m$, $\zeta$ and $N$
for which the model performs optimally. A univariate marginal distribution
algorithm \citep{Behera:03} was used to get near optimal parameters while fixing $N=400$ and $\hbar=1.0$.
The selected values of these parameters are as
follows for all levels of SNR:
\begin{equation}
\beta = 0.86;\;\;\; m = 2.5;\;\;\; \zeta = 2000;\;\;\; 
\end{equation}

The comparative performance of eye tracking in terms of rms error for all
the noise levels is
shown in Table I. It is easily seen from Table I that the rms tracking error
of RQNN is much less than that of the Kalman filter. Moreover, RQNN performs
equally well for all the three categories of noise levels, whereas the performance
of the Kalman filter
degrades with the increase in noise level. In this sense we can say that our model 
performs the tracking with a greater efficiency compared to the Kalman filter.
The exact nature of trajectory tracking
is shown for 0 dB SNR in FIG. \ref{fig:dc_track3}. In this figure, the noise
envelope is shown, and obviously its size is large due to a high noise
content in the signal. The figure shows the trajectory of the eye movement as the eye
focuses on a fixed
target. 
\begin{table}
\label{tab:ng}
\caption{Performance comparison between Kalman filter and RQNN for various levels of Gaussian noise}
\begin{tabular}{llcc}
\\\hline
Noise  level       &RMS  error             &RMS  error\\
in dB              &for  Kalman filter     &for  RQNN\\\hline
20                 &0.0018                 &0.000040\\
6                  &0.0270                 &0.000062\\
0                  &0.0880                 &0.000090\\\hline
\end{tabular}
\end{table}
To better appreciate the tracking performance, an error plot is
shown in FIG. \ref{fig:error}.  Although Kalman filter tracking is
continuous, the RQNN model tracking consists of 'jumps' and 'fixations'. As the
alignment of the eye pupil becomes closer to the target position, 
the 'fixation' time also increases. Similar tracking behaviour was also observed for the SNR values of 20 and 6 dB. These theoretical results are very
interesting when compared to experimental results in the field of
eye-tracking. In eye-tracking experiments, it is known that eye movements in
static scenes are not performed continuously, but consist of "jumps"
(saccades) and "rests" (fixations). Eye-tracking results are represented as
lists of fixation
data.  Furthermore, if the information is simple or
familiar, eye movement is comparatively smooth. If it is tricky or new, the eye
might pause or even flip back and forth between images. 
Similar results are given by our simulations. Our model tracks the dc signal which can
be thought of as equivalent to a static scene, in discrete steps rather than in a continuous
fashion. This is very clearly understood from the tracking error in FIG.
\ref{fig:error}. 

The other interesting aspect of the results is the movement of wave packets. In
Figure \ref{fig:dc_track3} (\textit{bottom}), snapshots of wave packets are plotted at different instances
corresponding to marker points as shown along the desired trajectory. It can be noticed that
a very flat initial Gaussian wave packet first moves to the left, and then  proceeds
toward the right until the mean of the wave packet exactly matches the
actual spatial
position. A similar pattern of movement of wave packets was also noticed in the case
of 20 and 6dB SNR. The wave packet movement is compared with our
previous work \citep{Behera:03} in FIG. \ref{fig:ln_rqnn}. The
initial wave packet in the previous model first splits into two
parts, then moves in a continuous fashion, 
ultimately going into a state with a mean of approximately 2 but with high
variance. In contrast, in the present model there is no splitting of the wave
packet, movement is discrete and variance is also much smaller. Thus the
soliton behavior of the present model is very much pronounced. 

\begin{figure}[htbp]
  \begin{center}
    \includegraphics[width=6.0cm,angle=0]{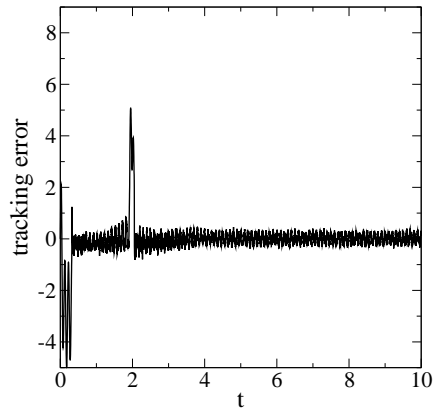}
    \caption{\small Saccadic and pursuit movement of eye during dynamic
      trajectory following } 
    \label{fig:error_sin}
  \end{center}
\end{figure}

\begin{figure}[htbp]
  \begin{center}
    \includegraphics[width=6.0cm,angle=0]{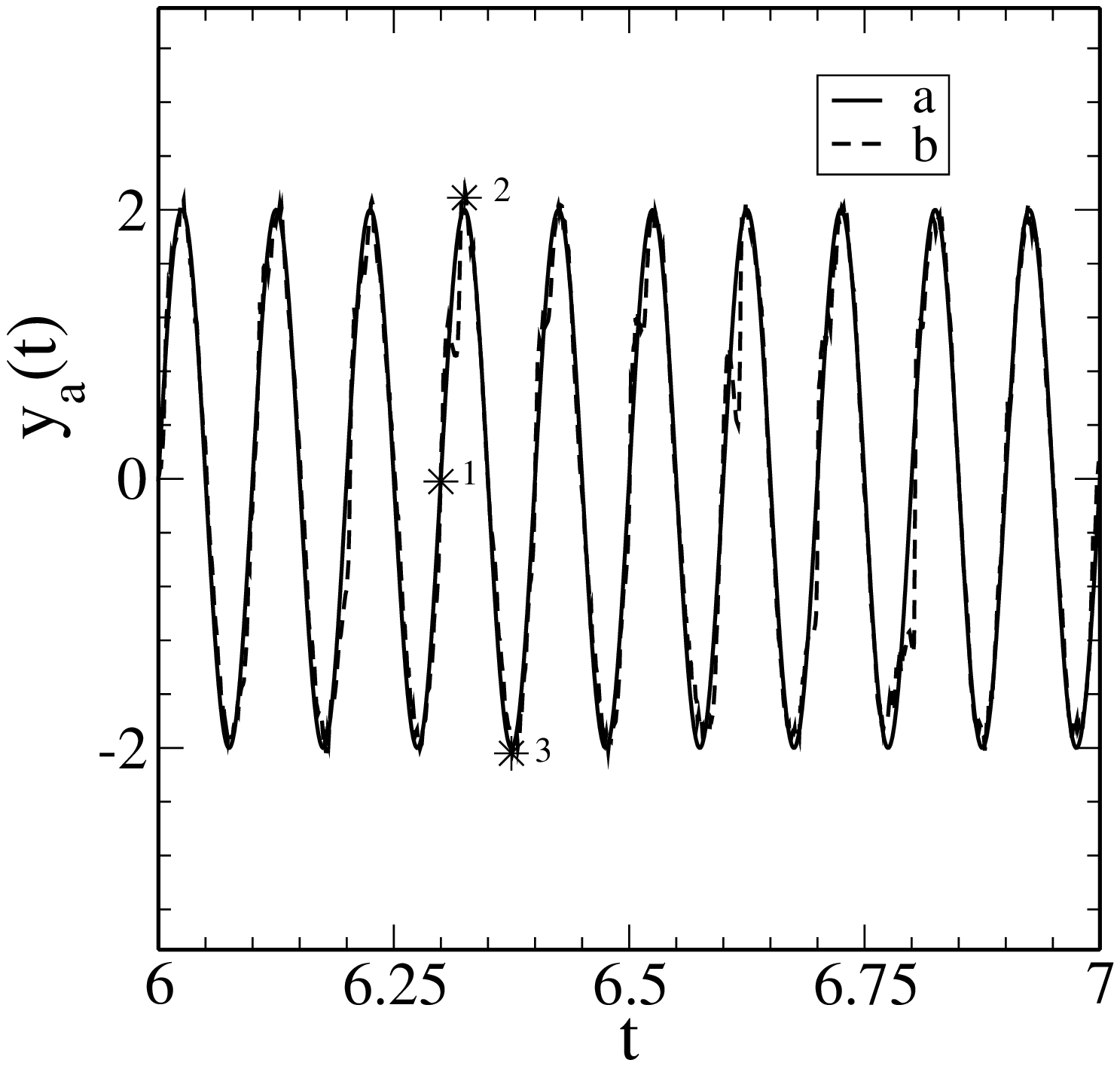} 
    \includegraphics[width=6.0cm,angle=0]{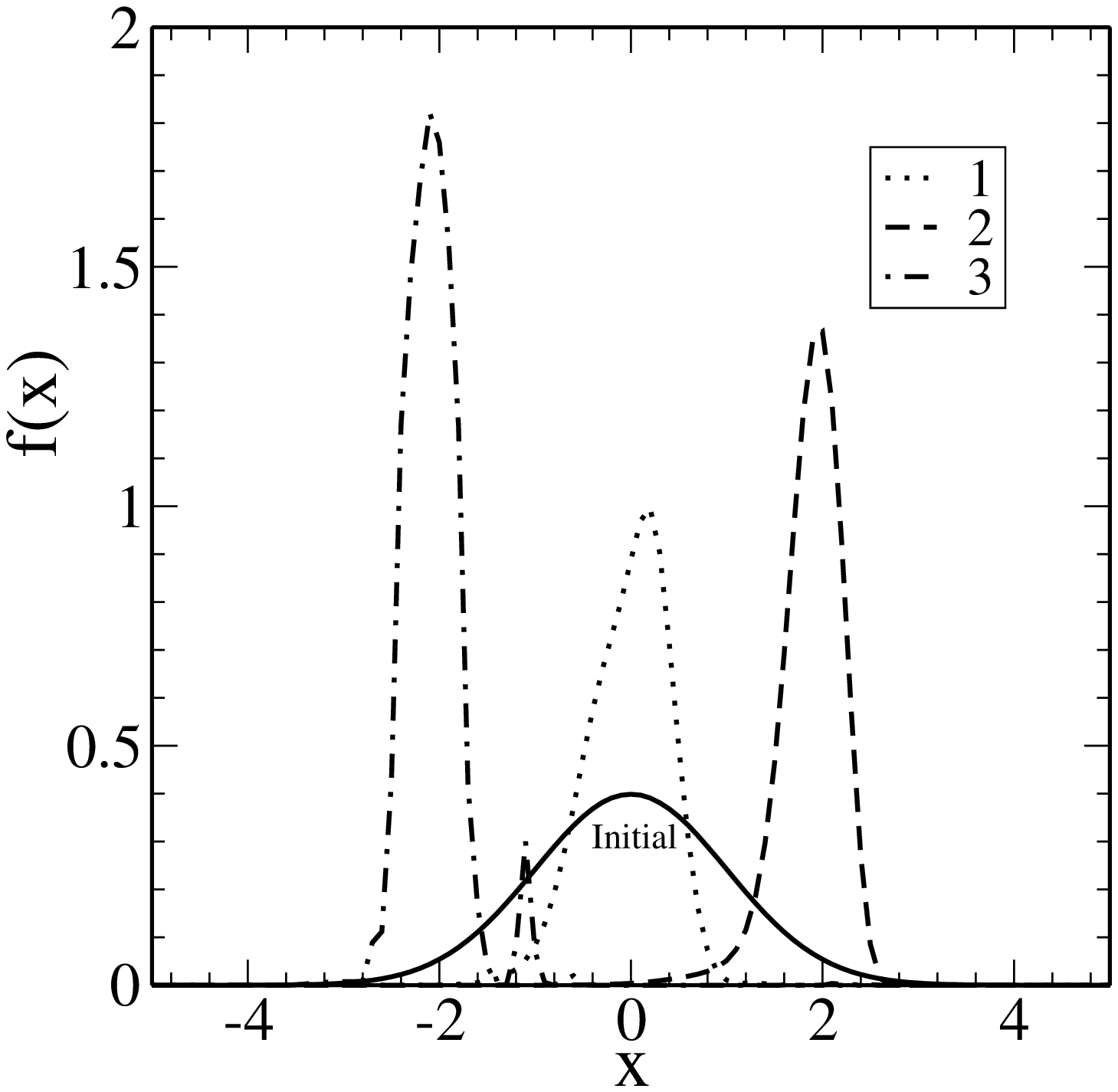}
    \caption{\small(\textit{top}) Eye tracking of a moving target in a noisy
      environment of 20dB SNR: 'a' respresents a moving target, 'b' represents
      target tracking using RQNN model; (\textit{bottom})  The snapshots of the
      wave packets at different instances corresponding to the marker points
      (1,2,3) as shown in the \textit{top}
      figure. The solid line represents
      the initial wave packet assigned to the Schroedinger wave equation.} 
    \label{fig:sin_track}
  \end{center}
\end{figure}

To analyze the eye movement following a moving target, a sinusoidal signal
$y_a(t)=2 sin 2\pi 10 t$ is taken as the desired dynamic trajectory. This signal is embedded 
in 20 dB Gaussian noise. The parameter values for tracking this signal were fixed 
at $\beta=0.01, m=1.75$ and $\zeta=-250$. It is observed that during the training phase, 
the wave packet jumps from time to time, thus changing the tracking error until 
a steady state  trajectory following is achieved. This feature is clearly 
understood from the tracking error plot which is shown in FIG. \ref{fig:error_sin}. In this figure, it is shown that the wave packet 
has jumped six times before the first smooth
movement started. Again, this jump took place four times before the second
smooth movement started and ultimately achieved a steady state. When the steady state is achieved, 
the tracking is efficient and the
wave packet movement is continuous, as shown in FIG. \ref{fig:sin_track}. 
The snapshots of the wave packets are plotted for three different instances of time indicated 
by the marker points (1,2,3) as shown in the trajectory tracking. When the signal is at position 1,
the corresponding wave packet has a mean  at 0. When the signal is at position 2,
the corresponding wave packet has a mean at +2, and the mean of the wave packet moves to
-2 when the signal goes to position 3. This type of continuous movement of
wave packet takes place after a series of 'jumps' of random nature. During
continuous movement of the wave packets, trajectory tracking is smooth, 
denoted as smooth pursuit movement    
in the context of biological eye tracking. This theoretical result is very similar in nature as what has been observed 
experimentally \citep{Bahill:80}.

\section{Conclusion}

 The nature of eye movement has been studied in this article using the proposed RQNN model, where the
 predictor and motor control are assumed to be ideal. The most important
 finding is that our theoretical model of eye-tracking agrees with previously observed experimental
 results. The model predicts that eye movements will be of
 saccadic type while following a static trajectory. In the case of dynamic
 trajectory following, eye movement consists of saccades and smooth pursuits. In this sense, the proposed quantum brain concept in this paper is very
 successful in explaining the nature of eye movements. Earlier explanation \cite{Bahill:79} for saccadic movement has been primarily attributed to motor control mechanism whereas the present model emphasizes that such eye movements are due to decision making process of the brain - albeit quantum brain. Thus the significant contribution of this paper to explain biological eye-movement as a neural information processing event may   inspire researchers to study quantum brain models from the biological perspective. 

 The other significant contribution of this paper is the prediction efficiency
 of the proposed model over the prevailing model. The
 stochastic filtering of a dc signal using RQNN is 1000 times more accurate
 compared to a Kalman filter.  

At this point the paper is silent about exact biological connection between classical and quantum brain since it is not clear to us. The model just assumes that the quantum brain is excited by the potential field set up by the classical brain. Another obvious question is that of decoherence. In this regard, we admit that the model proposed here is highly idealized since we have used Schroedinger wave equation. We intend to replace Schroedinger wave equation by density matrix approach in our future work.  Also, the phase transition analysis of closed form dynamics, given in equation \eqref{eq:wave1} with respect to various parameters $m,\zeta,\beta$ and $N$, has been kept for future work.

Finally, we believe that apart from the computational power derived from quantum computing, quantum learning
systems will also provide a potent framework to study the subjective aspects of
the nervous system \cite{Harald:04}. The challenge to bridge the gap between physical and mental (or
objective and subjective) notions of matter may be most successfully met
within the framework of quantum learning systems. In this framework, we have proposed a
notion of a quantum brain, and a Recurrent Quantum Neural Network has been
hypothesized as a first step towards a neural computing model.




\end{document}